\documentclass[reprint,amsmath,amssymb,superscriptaddress]{revtex4-2}

\usepackage{graphicx}
\usepackage{color}

\begin{document}

\title{Classical-quantum correspondence in the noise-based dissipative systems}

\author{Jiarui Zeng}
\affiliation{School of Physics and Optoelectronic Engineering, Hainan University, Haikou 570228, China}
\affiliation{Department of Physics, South China University of Technology, Guangzhou 510640, China}

\author{Guo-Hao Xu}
\affiliation{Department of Physics, South China University of Technology, Guangzhou 510640, China}

\author{Weijie Huang}
\affiliation{Department of Physics, South China University of Technology, Guangzhou 510640, China}

\author{Yao Yao}
\email{yaoyao2016@scut.edu.cn}
\affiliation{Department of Physics, South China University of Technology, Guangzhou 510640, China}
\affiliation{State Key Laboratory of Luminescent Materials and Devices, South China University of Technology, Guangzhou 510640, China}

\date{\today}

\begin{abstract}

We investigate the correspondence between classical noise and quantum environments. Although it has been known that the classical noise can be mapped to the quantum environments only for pure dephasing and infinite-temperature dissipation processes, we describe that this limitation can be circumvented by introducing auxiliary systems and conservation. Taking a two-level system as an example, we construct the so-called central spin model with its couplings fluctuating as the classical noise, and then acquire its statistical-average dynamics which captures the dissipations beyond the infinite temperature. By adjusting the number of the auxiliary systems and their initial states, the noise-based model reproduces both Markovian and non-Markovian evolutions. It is also found that different quantities of the two-level system are governed by different model parameters, indicating that the constructed model is an efficient simulator for specific observables, rather than an equivalent form of a realistic open system. In addition, the model is also applicable to investigate topical mechanisms of the open systems, e.g. negative temperatures and asymmetric equidistant quenches.

\end{abstract}

\maketitle

\section{Introduction}
\label{secIntroduction}

Classical noise has implicit but significant influences in a number of physical domains. Hitherto it has been applied into extensive investigations such as semi-classical dynamics \cite{tully1990molecular, barbatti2011nonadiabatic}, noise-assisted transports \cite{rebentrost2009environment, maier2019environment}, dynamical decoupling \cite{uys2009optimized, lo2014preserving, szankowski2017environmental}, etc. These studies inspire attention on the influences of the classical noise, because it seems to be exhibited as thermal environments of open systems. For example, it is found that the electric-field noise result a motional ion in anomalous heating in trapped-ion devices, as its vibrational number always increases with respect to time \cite{brownnutt2015ion, hite2012100, zeng2023theoretical}. In spin systems, the magnetic-field noise is also notorious because of its decoherence effects \cite{schmidt2003coherence, merkel2019magnetic}. In order to reduce the noise-induced decoherence, the spin should be frequently flipped to dynamically average out the spin-environment couplings \cite{viola1998dynamical, viola1999dynamical}.

Generally speaking, for a closed system coupled with the classical noise, its time-dependent evolutions should be always unitary and the decoherence of the system is out of context, because there are no external degrees of freedom. However, with statistical averaging over an ensemble, the decay can be observed in both the diagonal and off-diagonal elements in the density matrix of the system, indicating the decoherence occurs \cite{kiely2021exact}. Such the phenomenon is impossible in the closed system taking the noise out of account, unless focusing on the part of the system which is the basic idea of the eigenstate thermalization hypothesis \cite{deutsch1991quantum, srednicki1994chaos, d2016quantum, deutsch2018eigenstate, abanin2019colloquium}. In this sense, it seems that quantum environments, the origins of the decoherence effects such as bosonic and fermionic baths in the realistic open systems, may be substituted by the classical noise. Their interesting correspondence has been applied into quantum simulations, achieving numerous open systems such as the light-harvesting model, Bose--Hubbard model, and Ising chain \cite{mostame2012quantum, potovcnik2018studying, wang2018efficient, chenu2017quantum}.

Because there are still differences between the classical noise and quantum environments, a fundamental question inevitably arises: when it will work if describing the decoherence effects with the classical noise. Normally the decoherence includes pure dephasing and dissipations, of which the former one only involves the decay of the off-diagonal elements of the system density matrix, and the latter one further concerns transitions between quantum states of the system. It has been demonstrated that any pure dephasing can be described with the classical noise, given an artificial correlation function determined by the quantum environment \cite{gu2019can}. This insightful perspective is helpful for studies of the open systems with low-frequency environments \cite{bergli2009decoherence, yang2016quantum}, but remains the question of the dissipations unclear. Currently it is believed that because of the commutativity of the classical noise in its correlation function, one can not rebuild asymmetric processes of emissions and absorptions between the system states \cite{gu2019can, tanimura1989time}. This means that the quantum environments can be substituted by the classical noise only at an infinite temperature. In order to describe the finite-temperature dissipations, several strategies beyond the classical noise are proposed, such as propagating the systems in a non-unitary way \cite{burgarth2017can, haga2023quasiparticles} and considering phenomenological corrections \cite{bastida2007ehrenfest, aghtar2012juxtaposing}.

However, we notice that this limitation can be circumvented within the framework of the classical noise. The basic idea is to introduce auxiliary systems and conservation to construct effective quantum noise. In this paper, we take a two-level system as an example and construct a noise-based model to reproduce the dissipation dynamics at the finite temperatures. The time-dependent evolutions of system observables are in good agreement with that calculated from the quantum master equation, indicating the effectiveness of the model. Based on these results, we further discuss the correspondence between the classical noise and quantum environments. Surprisingly, it is found that the noise-based model is an efficient simulator rather than an equivalent form of the realistic open systems, because it provides correct descriptions only if focusing on the specific observables. The model captures the whole dissipation dynamics while considering an infinite number of the auxiliary systems, providing an exotic perspective on the classical-quantum correspondence. In addition, to further illustrate that the noise-based model captures the finite-temperature dissipations beyond the two-level systems, we discuss the possible applications for investigations of topical mechanisms of the open systems, including the negative temperatures and equidistant quenches.

The rest of this paper is organized as follows. We first describe the Lindblad master equation of the two-level systems in Sec.~\ref{secMaster}, which are our targets to be described with the classical noise. In Sec.~\ref{secModel1}, we explain why the previously proposed model can not rebuild the asymmetric processes of the absorptions and emissions, i.e. the finite-temperature dissipations. In order to circumvent this limitation, we introduce the auxiliary systems and conservation into the two-level systems, to construct the new noise-based model and obtain its corresponding dynamics in Sec.~\ref{secModel2}. We show both the diagonal and off-diagonal elements of the system density matrix are in accordance with the results of the master equation, through adjusting the number of the auxiliary systems and their initial states. The correspondence between the classical noise and quantum environments is also discussed based on these results. In Sec.~\ref{secBeyond}, we describe that the noise-based model can be extended beyond the two-level systems. Finally, Sec.~\ref{conclusions} draws a conclusion.

\section{Quantum master equation of open systems}
\label{secMaster}

Before describing the model for the finite-temperature dissipations with the classical noise, we first introduce a two-level system interacting with a quantum environment as an example. We assume that its reduced dynamics obeys the Lindblad master equation, which is written as (throughout this paper $\hbar = 1$) \cite{lindblad1976generators, manzano2020short}
\begin{equation}
  \dot{\rho}(t) = -i \left[ H_\mathrm{s}, \rho(t) \right] +\mathcal{D} \left[ \rho(t) \right].
  \label{eq_masterEquation}
\end{equation}
Herein, $H_\mathrm{s} = \omega_\mathrm{s} \sigma_z/2$ and $\rho(t)$ denote the Hamiltonian and density matrix of the system respectively, with $\sigma_z$ the spin-1/2 operator with $\sigma_z | \pm \rangle = \pm | \pm \rangle$. $\left[ A, B \right] = AB -BA$ is the commutator. Moreover, the term $\mathcal{D} \left[ \cdots \right]$ describes the dissipation effects resulting from the quantum environment:
\begin{align}
  \mathcal{D} \left[ \rho(t) \right] =\ &\Gamma_e \left( \sigma^- \rho(t) \sigma^+ -\frac{1}{2} \left\{ \sigma^+ \sigma^-, \rho(t) \right\} \right) \nonumber\\
  \ &+\Gamma_a \left( \sigma^+ \rho(t) \sigma^- -\frac{1}{2} \left\{ \sigma^- \sigma^+, \rho(t) \right\} \right),
  \label{eq_dissipationTerm}
\end{align}
where $\Gamma_e$ and $\Gamma_a$ are the emission and absorption rates, respectively. The latter rate can be also regarded as the stimulated emission rate and thus the spontaneous emission rate reads $\Gamma_0 = \Gamma_e -\Gamma_a$. $\left\{ A, B \right\} = AB +BA$ refers to the anti-commutator, and $\sigma^\pm$ denotes the flipping operator satisfying $\sigma^\pm | \mp \rangle = | \pm \rangle$ and $\sigma^\pm | \pm \rangle = 0$. For the bosonic baths at thermal equilibrium, the absorption and spontaneous emission rates possess the following relationship \cite{breuer2002theory}:
\begin{equation}
  \frac{\Gamma_0}{\Gamma_a} = \frac{1}{N(\omega_\mathrm{s})} = e^{\omega_\mathrm{s} /k_\mathrm{B} T} -1,
  \label{eq_relationship}
\end{equation}
where $N(\omega_\mathrm{s})$ denotes the Bose--Einstein distribution, $k_\mathrm{B}$ is the Boltzmann constant, and $T$ is an environment temperature. 

The master equation can be further traced to be
\begin{equation}
  \dot{\rho}_z(t) = \mathrm{Tr} \left[ \sigma_z \dot{\rho}(t) \right] = -\Gamma_0 -2\Gamma_d \rho_z(t)
  \label{eq_masterEquationz}
\end{equation}
and
\begin{equation}
  \dot{\rho}_+(t) = \mathrm{Tr} \left[ \sigma^+ \dot{\rho}(t) \right] = i\omega_\mathrm{s} \rho_+(t) -\Gamma_d \rho_+(t)
  \label{eq_masterEquationp}
\end{equation}
with $\Gamma_d = \left( \Gamma_e +\Gamma_a \right) /2$. Equations~(\ref{eq_masterEquationz}) and (\ref{eq_masterEquationp}) describe the time-dependent evolutions of different observables, which are our targets described with the noise.

\section{Classical noise for infinite-temperature dissipations}
\label{secModel1}

Herein, we describe why the previously proposed model can not describe the dissipations unless considering an infinite temperature \cite{gu2019can}. The Hamiltonian of the model reads
\begin{equation}
  H(t) = H_\mathrm{s} +\eta(t) \sigma^+ +\eta^*(t) \sigma^-,
  \label{eq_model1}
\end{equation}
where $\eta(t)$ is the classical noise with $\overline{\eta^*(t') \eta(t)} = \gamma(t',t)$ and $\overline{\eta(t') \eta(t)} = \overline{\eta(t)} = 0$. The overline denotes the statistical averaging over all stochastic processes. Considering the white noise $\gamma(t',t) = \gamma \delta(t'-t)$, with $\delta(t'-t)$ the Dirac function, the von Neumann equation can be written as
\begin{equation}
  \dot{\rho}(t) = -i\left[ H_\mathrm{s} +\eta(t) \sigma^+ +\eta^*(t) \sigma^-, \rho(t) \right].
\end{equation}
Through iterative substitution and statistical averaging, we then obtain
\begin{align}
  \dot{\rho}(t) =\ &-i \left[ H_\mathrm{s}, \rho(t) \right] -\int_0^{t} \mathrm{d} \tau \, \overline{\eta(t) \eta^*(\tau)} \left[ \sigma^+, \left[ \sigma^-, \rho(\tau) \right] \right] \nonumber\\
  \ &-\int_0^t \mathrm{d} \tau \, \overline{\eta^*(t) \eta(\tau)} \left[ \sigma^-, \left[ \sigma^+, \rho(\tau) \right] \right] \nonumber\\
  =\ &-i \left[ H_\mathrm{s}, \rho(t) \right] +\mathcal{D} \left[ \rho(t) \right]
\end{align}
with the dissipation term
\begin{align}
  \mathcal{D} \left[ \rho(t) \right] =\ &-\gamma \left[ \sigma^+, \left[ \sigma^-, \rho(t) \right] \right] -\gamma \left[ \sigma^-, \left[ \sigma^+, \rho(t) \right] \right] \nonumber\\
  =\ &2\gamma \left( \sigma^- \rho(t) \sigma^+ -\frac{1}{2} \left\{ \sigma^+ \sigma^-, \rho(t) \right\} \right. \nonumber\\
  \ &\left. +\sigma^+ \rho(t) \sigma^- -\frac{1}{2} \left\{ \sigma^- \sigma^+, \rho(t) \right\} \right).
  \label{eq_dissipationTerm1}
\end{align}
Equation~(\ref{eq_dissipationTerm1}) leads the time-dependent expectations of the system operators to
\begin{equation}
  \dot{\rho}_z(t) = -4\gamma \rho_z(t)
  \label{eq_masterEquationz1}
\end{equation}
and
\begin{equation}
  \dot{\rho}_+(t) = i\omega_\mathrm{s} \rho_+(t) -2\gamma \rho_+(t).
\end{equation}
Equation~(\ref{eq_masterEquationz1}) indicates an inherent difficulty of describing the transitions between the system states, because the process of the spontaneous emission is omitted. Actually, the model corresponds to the realistic open system only when $\Gamma_e = \Gamma_a = 2\gamma$, i.e. only when the temperature $T \rightarrow \infty$ according to Eq.~(\ref{eq_relationship}).

\section{Noise-based model coupled with auxiliary systems}
\label{secModel2}

\subsection{Methodology}

Although the dissipations at the finite temperatures can not be described with Eq.~(\ref{eq_dissipationTerm1}), it is noticed that this limitation results from the commutativity of the classical noise, $\overline{\eta(t) \eta^*(\tau)} = \overline{\eta^*(\tau) \eta(t)}$, and can be circumvented. The basic idea is to introduce the auxiliary two-level systems with their flipping operators $[s^+, s^-] \neq 0$. By inserting $s^\pm$ into Eq.~(\ref{eq_dissipationTerm1}), we then construct artificial non-commutativity $\overline{\eta(t) s^- \eta^*(\tau) s^+} \neq \overline{\eta^*(\tau) s^+ \eta(t) s^-}$, and therefore the spontaneous emissions. However, the expectations of the auxiliary operators might cause the emission and absorption rates to be uncertain. We further introduce the conservation to map the auxiliary operators $s^\pm s^\mp$ to the system operators $\sigma^\pm \sigma^\mp$, and utilize the algebraic properties of the flipping operators of the two-level systems to cancel them.

\begin{figure}
  \includegraphics[width=\linewidth]{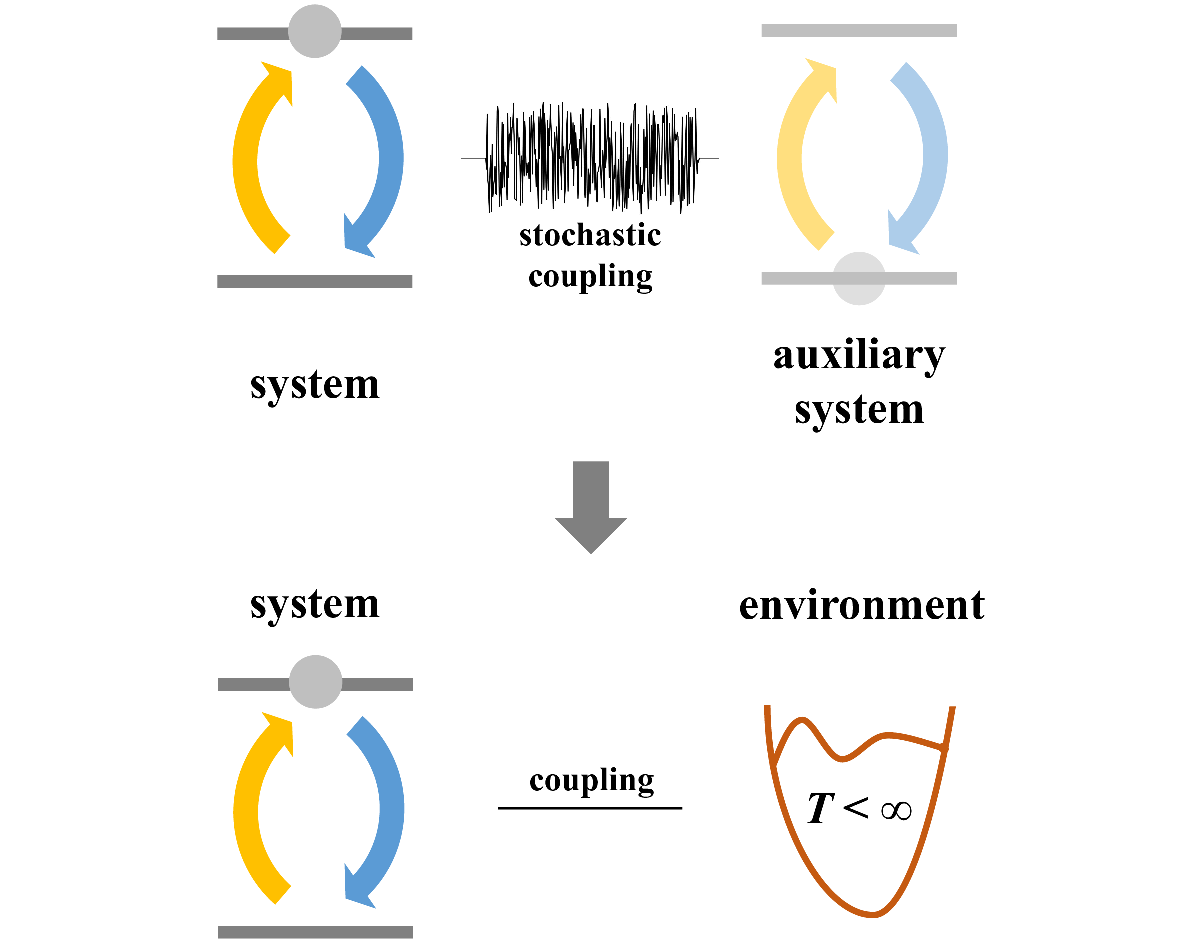}
  \caption{Schematic of the noise-based model considering auxiliary two-level systems.}
  \label{fig_schematic}
\end{figure}

Specifically, we here introduce $N$ auxiliary two-level systems, leading to the central spin model
\begin{equation}
  H(t) = H_\mathrm{s} +\sum_{i = 1}^N \left( \eta_i(t) s_i^- \sigma^+ +\eta_i^*(t) s_i^+ \sigma^- \right)
  \label{eq_model2}
\end{equation}
as illustrated in Fig.~\ref{fig_schematic}, where $s_i^\pm$ is the flipping operator of the $i$th auxiliary system. Then the model holds the conservation of the total spin, as $\sigma^+ \sigma^- +\sum_{i = 1}^N s_i^+ s_i^- = N_+$ and $\sigma^- \sigma^+ +\sum_{i = 1}^N s_i^- s_i^+ = N_-$, with $N_+ +N_- = N +1$. All the noise has $\overline{\eta_i^*(t') \eta_i(t)} = \gamma(t',t)$, $\overline{\eta_i(t') \eta_i(t)} = \overline{\eta_i(t)} = 0$, and $\overline{\eta_i^*(t') \eta_j(t)} = \overline{\eta_i(t') \eta_j(t)} = 0$ ($i, j$ from $1$ to $N$ and $i\neq j$). Similar to the previous section, we take the white noise into account, and obtain
\begin{align}
  \ &\mathcal{D} \left[ \rho(t) \right] \nonumber\\
  =\ &2\gamma \sum_{i = 1}^N \left( s_i^+ \sigma^- \rho(t) s_i^- \sigma^+ -\frac{1}{2} \left\{ s_i^- s_i^+ \sigma^+ \sigma^-, \rho(t) \right\} \right. \nonumber\\
  &\left. +s_i^-\sigma^+ \rho(t) s_i^+ \sigma^- -\frac{1}{2} \left\{ s_i^+ s_i^- \sigma^- \sigma^+, \rho(t) \right\} \right).
\end{align}
It is found that the auxiliary operators have been inserted into the dissipation term. If we only focus on the specific physical quantity $A$ of the realistic system, the dissipation term can be transformed to have an effective form:
\begin{align}
  \ &\mathrm{Tr} \left[ A \mathcal{D} \left[ \rho(t) \right] \right] \nonumber\\
  =\ &2\gamma \sum_{i = 1}^N \mathrm{Tr} \left[ s_i^- s_i^+ \left( \sigma^+ A \sigma^- -\frac{1}{2} \left\{ \sigma^+ \sigma^-, A \right\} \right) \rho(t) \right. \nonumber\\
  \ &\left. +s_i^+ s_i^- \left( \sigma^- A \sigma^+ -\frac{1}{2} \left\{ \sigma^- \sigma^+, A \right\} \right) \rho(t) \right] \nonumber\\
  =\ &2\gamma \mathrm{Tr} \left[ N_-^A \left( \sigma^+ A \sigma^- -\frac{1}{2} \left\{ \sigma^+ \sigma^-, A \right\} \right) \rho(t) \right. \nonumber\\
  \ &\left. +N_+^A \left( \sigma^- A \sigma^+ -\frac{1}{2} \left\{ \sigma^- \sigma^+, A \right\} \right) \rho(t) \right],
  \label{eq_dissipationTerm2}
\end{align}
where $N_-^A = N_- -\sigma^- \sigma^+$ and $N_+^A = N_+ -\sigma^+ \sigma^-$. $N_\pm^A$ equals to $N_\pm$ for $A = \sigma_z$, and can be arbitrary number under the constraint $N_-^A +N_+^A = N_- +N_+ -1 = N$ for $A = \sigma^+$. Equation~(\ref{eq_dissipationTerm2}) lead the dissipation dynamics to
\begin{equation}
  \dot{\rho}_z(t) = -2\gamma \left( N +1 -2N_+ \right) -2\gamma \left( N +1 \right) \rho_z(t)
  \label{eq_masterEquationz2}
\end{equation}
and
\begin{equation}
  \dot{\rho}_+(t) = i\omega_\mathrm{s} \rho_+(t) -\gamma N \rho_+(t),
  \label{eq_masterEquationp2}
\end{equation}
respectively.

\subsection{Discussions}

\begin{figure}
  \includegraphics[width=\linewidth]{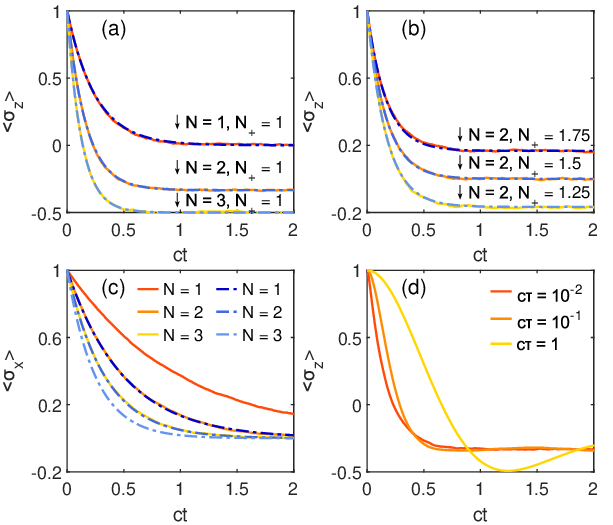}
  \caption{$\langle \sigma_z \rangle$ calculated from the noise-based model (warm-color solid lines) and Lindblad master equation (cool-color dash-dotted lines) by adjusting (a) the number of the auxiliary systems $N$ and (b) their initial states referring to $N_+$. (c) $\langle \sigma_x \rangle$ and $\langle \sigma_z \rangle$ calculated from the noise-based model (warm-color solid lines) and Lindblad master equation (cool-color dash-dotted lines) by adjusting $N$ and memory time $c\tau$, respectively. Other parameters are $\omega_\mathrm{s}/c = 0.1$ and $c\Delta t = 10^{-4}$. In panel [(a)--(c)], the memory time is $c\tau = 10^{-3}$. In panel (c), $N_+ = 1$. In panel (d), $N = 2$ and $N_+ = 1$. Computational details can be also found in appendix \ref{secComputational}.}
  \label{fig_calculated}
\end{figure}

To show the performance of the noise-based model, we conduct numerical calculations with their computational details described in appendix \ref{secComputational}. Taking the white noise into account, we find the model well rebuilds the dissipations for $\Gamma_e \neq \Gamma_a$, as its results are in agreement with that acquired from the master equation in Figs.~\ref{fig_calculated}(a) and \ref{fig_calculated}(b). There are two ways to prepare the model parameters determined by the realistic open system: to adjust the number of the auxiliary systems or their initial states. The former one is exhibited in Fig.~\ref{fig_calculated}(a), with the initial conditions adopted as $|\psi(0) \rangle = |+\rangle \otimes_{i=1}^{N} |-\rangle_i$, resulting in $\Gamma_e = 2\gamma N$ and $\Gamma_a = 2\gamma$. If there is only one auxiliary system with its initial state orthogonal to the counterpart of the central system, the model becomes equivalent with Eq.~(\ref{eq_model1}), and the corresponding time-dependent evolutions reduce to the infinite-temperature dissipations due to the equal emission and absorption rates. In order to rebuild the finite-temperature dissipations, the number of the auxiliary systems should be $N > 1$. In addition, we can further flip part of the auxiliary systems, leading to the more flexible model parameters $\Gamma_e = 2\gamma N_-$ and $\Gamma_a = 2\gamma N_+$.

If the initial conditions of the auxiliary systems are not polarized in the $z$ direction, the model parameters would approach a more extensive regime. In Fig.~\ref{fig_calculated}(b), we fix the number of the auxiliary systems $N$ and adjust their initial conditions, leading to the continuously changing $N_\pm$ and corresponding results. For each evolution, the initial wavefunction of the system is not unique, because the majority of the information is neglected in the statistical averaging and thus the dissipation dynamics is determined by a few parameters, i.e. $N_\pm$ in Fig.~\ref{fig_calculated}(b). It should be also mentioned for $N_+ = 1.75$ that the model parameters result in an unphysical regime, as the population at the higher level is larger than that at the lower one at the steady state. Nevertheless, the result indicates the versatility of the noise-based model which is able to describe extensive scenarios.

The above results manifest that the dissipations at the finite temperatures can be well rebuilt with the classical noise, as the model predicts the reliable results for the transitions between the system states. While focusing on the off-diagonal elements of $\rho(t)$, it is found that they have deviations despite capturing the decoherence behaviors. In Fig.~\ref{fig_calculated}(c), we depict $\langle \sigma_x \rangle$ acquired from the noise-based model and master equation, with the same model parameters as that in Fig.~\ref{fig_calculated}(a) and the central system propagating from $\left( |+\rangle +|-\rangle \right)/\sqrt{2}$. It can be clearly seen that lines obtained from the model with $N$ auxiliary systems always correspond to the master equation with the parameter $N-1$. This phenomenon can be understood by different $\Gamma_d$ in Eqs.~(\ref{eq_masterEquationz2}) and (\ref{eq_masterEquationp2}), meaning that the noise-based model can not provide the correct dynamics of both the diagonal and off-diagonal elements of $\rho(t)$ in the same calculation. In this context, the current model is not an equivalent form of a realistic open system, but an efficient simulator for specific observables. We also note for $N \rightarrow \infty$ that the model correctly captures the whole dissipation dynamics, and the auxiliary systems approach the realistic spin baths. Therefore, we conclude that the classical noise can be mapped to the finite-temperature environments only if focusing on the specific observables, namely part of the physical quantities. Nevertheless, the noise-based model paves an efficient path to describe the finite-temperature dissipations with the classical noise, and a perspective to view its role.

The non-Markovian evolutions can be also described utilizing the colored noise \cite{chenu2017quantum, costa2017enabling}. We change the memory time $c\tau$ of the classical noise and plot their corresponding results in Fig.~\ref{fig_calculated}(d), in which a crossover from incoherent decays to coherent oscillations can be observed. For the small memory time, $\langle \sigma_z \rangle$ is exhibited as exponential decays, resembling to the results in Fig.~\ref{fig_calculated}(a). While further increasing the memory time, the dynamics oscillating in underdamped sinusoidal behaviors for $c\tau = 10^{-1}$ and $1$, which is clearer for the latter case. Such the coherent oscillations refer to the non-Markovianity because of their information backflow from the environment to the system \cite{breuer2012foundations, wenderoth2021non, zeng2024perturbative}. Therefore, the noise-based model is applicable to investigate both Markovian and non-Markovian dynamics.

It is also an interesting question whether the noise used in Eq.~(\ref{eq_model2}) possesses a classical or quantum nature. Although the couplings in the noise-based model are generated from the classical noise, they combine with the auxiliary systems and form the effective quantum noise to gain the non-commutative property. In this sense, we consider the noise has the classical nature, but the model circumvents its limitation rather than breaks it. On the other hand, it might be confusing whether the auxiliary systems are equal to an environment, but actually the number of the auxiliary systems can be finite and even as few as one, and the whole system is normally regarded as the closed system.

\section{Applications}
\label{secBeyond}

The correspondence between the classical noise in the constructed model and quantum environments has been discussed. We here further describe that the noise-based model is able to capture the finite-temperature dissipative mechanisms beyond the two-level systems, through applications into two mechanisms, negative temperatures and asymmetric equidistant quenches. The former refers to inversion of level populations \cite{purcell1951nuclear}, which has many interesting properties such as heat flows from the negative-temperature to a positive-temperature environment \cite{ramsey1956thermodynamics}. For the latter one, it is found that thermal relaxations at the ambient temperatures are asymmetric from a cooler and hotter initial conditions, resulting in the so-called asymmetric equidistant quenches \cite{lapolla2020faster}. Depending on the model parameters, the quench dynamics exhibits quantum phase transitions \cite{manikandan2021equidistant}. Both of the two mechanisms can be realized with a three-level system consisting of different quantum environments. The system Hamiltonian is adopted as $H_\mathrm{s} = \sum_{n=\{1,2,3\}} E_n |n\rangle \langle n|$ and the dissipation term is 
\begin{align}
  \mathcal{D} \left[ \rho(t) \right] = \sum_{p\neq q} \Gamma_{pq} \left( L_{pq} \rho(t) L_{pq}^\dag -\frac{1}{2} \left\{ L_{pq}^\dag L_{pq}, \rho(t) \right\} \right),
\end{align}
where $L_{pq} = |p\rangle \langle q|$ is the transition operator with $p, q=\{1,2,3\}$.

\subsection{Negative temperatures}

We first consider a negative-temperature case, with the level energies of the model are $E_1 = 0$, $E_2 = E_H -E_C$, and $E_3 = E_H$. For the dissipation term, $\Gamma_{13} = \Gamma_H (N_H +1)$, $\Gamma_{31} = \Gamma_H N_H$, $\Gamma_{23} = \Gamma_C (N_C +1)$, and $\Gamma_{32} = \Gamma_C N_C$, where $N_{X=\{H,C\}} = 1/(\exp(\beta E_X) -1)$ and $\beta$ is an inverse temperature. Transitions between $|1\rangle$ and $|2\rangle$ are taken out of account as $\Gamma_{12} = \Gamma_{21} = 0$. Through these model parameters, the population inversion occurs at $|1\rangle$ and $|2\rangle$, leading to the so-called negative temperatures \cite{bera2024steady}.

To describe the mechanism and its resultant effect, it is practical to construct the noise-based model as $H(t) = H_\mathrm{s} +\sum_{X = \{H,C\}} \sum_{i = 1}^{M_X} \left( \eta_{X,i}(t) s_{X,i}^- L_X^\dag +\eta_{X,i}^*(t) s_{X,i}^+ L_X \right)$, with $L_H = L_{13}$, $L_C = L_{23}$, and the white-noise processes $\overline{\eta_{X,i}^*(t') \eta_{X,i}(t)} = \gamma_X \delta(t'-t)$. It is known that the model possesses the conservation $\sum_i^N s_{X,i}^- s_{X,i}^+ +L_X L_X^\dag = m_X$, where $m_X$ ($X = \{H,C\}$) is a constant. Then the effective dissipation term can be written as
\begin{align}
  \ &\mathrm{Tr} \left[ A \mathcal{D} \left[ \rho(t) \right] \right] \nonumber\\
  =\ &2\sum_{X} \gamma_X \mathrm{Tr} \left[ M_{-,X}^A \left( L_X^\dag A L_X -\frac{1}{2} \left\{ L_X^\dag L_X, A \right\} \right) \rho(t) \right. \nonumber\\
  \ &\left. +M_{+,X}^A \left( L_X A L_X^\dag -\frac{1}{2} \left\{ L_X L_X^\dag, A \right\} \right) \rho(t) \right].
\end{align}
Herein, $M_{-,X}^A = m_X -L_X L_X^\dag$, $M_{+,X}^A = M_X -m_X +1 -L_Y L_Y^\dag -L_Y^\dag L_Y$, with $Y = \{C,H,\ \mathrm{if}\ X = H,C\}$. In the negative-temperature systems, the interest is the population inversion and thus we concentrate on the diagonal elements of $\rho(t)$. Then both of $L_X L_X^\dag$ and $L_Y L_Y^\dag$ and their Hermitian conjugates are commutative with the observable $A$, leading to $\Gamma_X (N_X +1) = 2\gamma_X m_X$ and $\Gamma_X N_X = 2\gamma_X (M_X -m_X +1)$.

\subsection{Equidistant quenches}

The second mechanism is the asymmetric equidistant quenches. The energy of the lowest level is set as $E_1 = 0$ and the other two energies are variable. Environment-induced transitions between all pairs of the levels are allowed, with $\Gamma_{12} = \Gamma_{21} e^{\beta E_2}$, $\Gamma_{23} = \Gamma_{32} e^{\beta (E_3-E_2)}$, and $\Gamma_{13} = \Gamma_{31} e^{\beta E_3}$. By adjusting $\Gamma_{13}/\Gamma_{12}$ and $\Gamma_{23}/\Gamma_{12}$, the quantum phase transitions occur exhibited as the faster uphill or downhill relaxation \cite{manikandan2021equidistant}. The noise-based model is constructed by introducing three-level auxiliary systems:
\begin{equation}
  H(t) = H_\mathrm{s} +\sum_{p>q} \sum_{i = 1}^{M_{pq}} \left( \eta_{i,pq}(t) \tilde{L}_{i,pq} L_{pq} +\eta_{i,pq}^*(t) \tilde{L}_{i,pq}^\dag L_{pq}^\dag \right),
\end{equation}
with $\tilde{L}_{i,pq} = |p\rangle \langle q|_i$ referring to the $i$th auxiliary system and the classical noise $\overline{\eta_{i,pq}^*(t') \eta_{i,pq}(t)} = \gamma_{pq} \delta(t'-t)$. Then we have the conservation $\sum_i \tilde{L}_{i,pq} \tilde{L}_{i,pq}^\dag -L_{pq} L_{pq}^\dag = m_p$. Herein, $m_p$ is a constant. Through similar treatments of the previous subsection, the effective dissipation term is
\begin{align}
  \ &\mathrm{Tr} \left[ A \mathcal{D} \left[ \rho(t) \right] \right] \nonumber\\
  =\ &2\sum_{p>q} \gamma_{pq} \mathrm{Tr} \left[ M_{-,pq}^A \left( L_{pq}^\dag A L_{pq} -\frac{1}{2} \left\{ L_{pq}^\dag L_{pq}, A \right\} \right) \rho(t) \right. \nonumber\\
  \ &\left. +M_{+,pq}^A \left( L_{pq} A L_{pq}^\dag -\frac{1}{2} \left\{ L_{pq} L_{pq}^\dag, A \right\} \right) \rho(t) \right].
\end{align}
where $M_{-,pq}^A = m_q +L_{pq}^\dag L_{pq}$ and $M_{+,pq}^A = m_p +L_{pq} L_{pq}^\dag$. Assuming initial thermal conditions, the coherence can be neglected in the time-dependent evolutions \cite{manikandan2021equidistant}. In this context, the model parameters are adopted as $\Gamma_{12} = 2\gamma_{21} (m_1 +1)$, $\Gamma_{21} = 2\gamma_{21} (m_2 +1)$, $\Gamma_{23} = 2\gamma_{32} (m_2 +1)$, $\Gamma_{32} = 2\gamma_{32} (m_3 +1)$, $\Gamma_{13} = 2\gamma_{31} (m_1 +1)$, and $\Gamma_{31} = 2\gamma_{31} (m_3 +1)$.

\section{Conclusions}
\label{conclusions}

In this paper, we have investigated the correspondence between the classical noise and quantum environments within the framework of the Lindblad master equation. By introducing the auxiliary systems to the two-level system and holding the conservation, we construct the central spin model with its couplings fluctuating as the classical noise. The model well rebuilds the dissipations at the finite temperatures, which are in accordance with that calculated from the master equation. To acquire the designative emission and absorption rates, one can adjust both the number of the auxiliary systems and their initial states. Our results show that the limitation of the classical noise can be circumvented, and the noise can be mapped to the quantum environments beyond the infinite temperature. The correspondence is established only if focusing on the specific observations, which is attributed to the finite auxiliary systems. When the number of the auxiliary systems increases to infinity, the noise-based model approaches an open system and captures the whole dynamics for all the physical quantities. Based on these results, we discuss the potential applications to describe the finite-temperature dissipations beyond the two-level systems. We describe how the negative temperatures and asymmetric equidistant quenches can be realized with the classical noise.

The noise-based model may be also applicable for other physical and chemical schemes. For the electronic transitions in quantum dots or molecular systems and spin flipping, they often involve the interactions with the external environments such as fluctuating sources in solid states and solutions. The induced dissipations can be described with the semi-classical dynamics, e.g. the trajectory surface hopping methods \cite{chen2016accuracy}, which are considered to be weak in describing the low temperatures and coherence. It is possible to improve performance of the semi-classical dynamics with the noise-based model. In addition, the model may be realized in the quantum simulations such as superconducting circuits \cite{zagoskin2008superconducting}, achieving numerous open systems and corresponding investigations. However, it requires hard treatments for many degrees of freedom, because the number of the auxiliary systems and their couplings would be high-complexity. Nevertheless, we consider our efforts provide a new perspective on the classical noise, which is helpful for further studies.

\begin{acknowledgments}

This work was supported by the National Natural Science Foundation of China (Grants No. 12374107 and No. 11974118).

\end{acknowledgments}

\begin{appendix}

\section{Computational details}
\label{secComputational}

\begin{figure}
  \includegraphics[width=\linewidth]{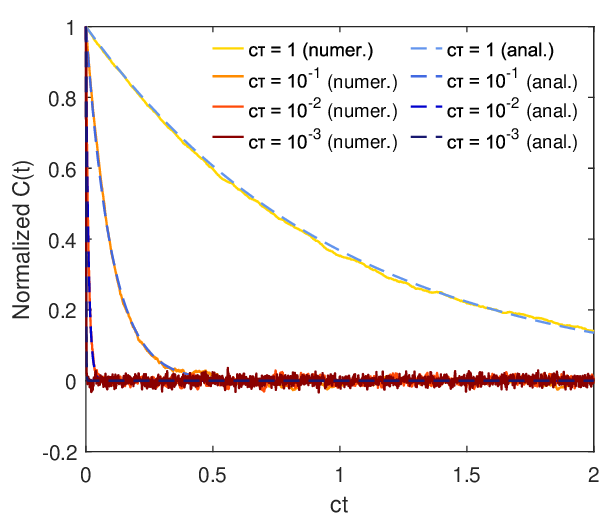}
  \caption{Normalized correlation function of the Johnson noise with $c\tau = 1$, $10^{-1}$, $10^{-2}$, and $10^{-3}$ (warm-color solid lines), averaging over $10^{4}$ noise processes. They are compared with the analytical results $\exp\left( -t/\tau \right)$ (cool-color dashed lines).}
  \label{fig_correlation}
\end{figure}

To calculate the numerical results of the noise-based model, we utilize the Johnson noise with the following spectrum:
\begin{equation}
  S(\omega) \sim \frac{1}{\omega^2 \tau^2 +1},
  \label{eq_noiseSpectrum}
\end{equation}
which allows us to obtain both white noise and colored noise by changing memory time $\tau$. For the very small $\tau$, the noise spectrum approaches the white noise with $S(\omega) \sim 1$. While increasing $\tau$, the noise exhibits itself as the typical Johnson noise, and further approaching the so-called Brownian noise $S(\omega) \sim \omega^{-2}$. To acquire time-dependent noise processes while performing the numerical calculations, an exact updating expression is employed \cite{gillespie1996mathematics, gillespie1996exact}:
\begin{equation}
  \xi(t+\Delta t) = \xi(t) e^{-\Delta t/\tau} +\left[ \frac{1}{c\tau} \left( 1 -e^{-2\Delta t/\tau} \right) \right]^{1/2} \chi(t),
  \label{eq_noise}
\end{equation}
and the complex noise can be therefore calculated with $\eta(t) = \xi_1(t) +i\xi_2(t)$. Herein, $c$ is the diffusion constant in the Ornstein--Uhlenbeck process, $\Delta t$ denotes the time step in the numerical calculations, and $\chi(t)$ refers to the normalized and unbiased white noise which can be generated with computer commands \cite{gillespie1996exact}, i.e. $\overline{\chi^2(t)} = 1$ and $\overline{\chi(t)} = 0$. Equation~(\ref{eq_noise}) generates the noise processes which are independent on the time step, avoiding a $\Delta t$-dependence problem if directly using $\chi(t)$. Moreover, $\xi(t)$ possesses the statistical average and correlation function as $\overline{\xi(t)} = \overline{\xi(0)} e^{-t/\tau}$ and
\begin{align}
  C(t',t) =\ &\overline{\xi(t') \xi(t)} = \overline{\xi^2(0)} e^{-(t'+t)/\tau} \nonumber\\
  \ &+\frac{1}{c\tau} \left( e^{-|t'-t|/\tau} -e^{-(t'+t)/\tau} \right),
\end{align}
respectively. We can acquire the correlation function of the exponential decay using an initial condition $\xi(0) = \left( 1/c\tau \right)^{-1/2} \, \chi(t)$, corresponding to the spectrum of the Johnson noise. In addition, other noise such as $\omega^{-1}$ noise can be also generated by summing up the Johnson noise with a series of the memory time $\tau$.

While performing the numerical calculations, the level difference and time step are chosen as $\omega_\mathrm{s}/c = 0.1$ and $c\, \Delta t = 10^{-4}$, respectively. The memory time is adopted as $c\tau = 10^{-3}$ for the white noise and $\{10^{-2}, 10^{-1}, 1\}$ for the colored noise. Their respective correlation functions $C(t) \equiv C(t,0)$ are illustrated in Fig.~\ref{fig_correlation}, in good agreement with the analytical results $\sim \exp(-t/\tau)$. We conduct the numerical calculations for $10^{4}$ times and then average their results. The final results are compared with that acquired from the Lindblad master equation Eq.~(\ref{eq_masterEquation}), given the parameters $\Gamma_e = 2N_-$ and $\Gamma_a = 2N_+$.

\end{appendix}

\bibliography{reference}

\begin{thebibliography}{48}%
\makeatletter
\providecommand \@ifxundefined [1]{%
 \@ifx{#1\undefined}
}%
\providecommand \@ifnum [1]{%
 \ifnum #1\expandafter \@firstoftwo
 \else \expandafter \@secondoftwo
 \fi
}%
\providecommand \@ifx [1]{%
 \ifx #1\expandafter \@firstoftwo
 \else \expandafter \@secondoftwo
 \fi
}%
\providecommand \natexlab [1]{#1}%
\providecommand \enquote  [1]{``#1''}%
\providecommand \bibnamefont  [1]{#1}%
\providecommand \bibfnamefont [1]{#1}%
\providecommand \citenamefont [1]{#1}%
\providecommand \href@noop [0]{\@secondoftwo}%
\providecommand \href [0]{\begingroup \@sanitize@url \@href}%
\providecommand \@href[1]{\@@startlink{#1}\@@href}%
\providecommand \@@href[1]{\endgroup#1\@@endlink}%
\providecommand \@sanitize@url [0]{\catcode `\\12\catcode `\$12\catcode
  `\&12\catcode `\#12\catcode `\^12\catcode `\_12\catcode `\%12\relax}%
\providecommand \@@startlink[1]{}%
\providecommand \@@endlink[0]{}%
\providecommand \url  [0]{\begingroup\@sanitize@url \@url }%
\providecommand \@url [1]{\endgroup\@href {#1}{\urlprefix }}%
\providecommand \urlprefix  [0]{URL }%
\providecommand \Eprint [0]{\href }%
\providecommand \doibase [0]{https://doi.org/}%
\providecommand \selectlanguage [0]{\@gobble}%
\providecommand \bibinfo  [0]{\@secondoftwo}%
\providecommand \bibfield  [0]{\@secondoftwo}%
\providecommand \translation [1]{[#1]}%
\providecommand \BibitemOpen [0]{}%
\providecommand \bibitemStop [0]{}%
\providecommand \bibitemNoStop [0]{.\EOS\space}%
\providecommand \EOS [0]{\spacefactor3000\relax}%
\providecommand \BibitemShut  [1]{\csname bibitem#1\endcsname}%
\let\auto@bib@innerbib\@empty
\bibitem [{\citenamefont {Tully}(1990)}]{tully1990molecular}%
  \BibitemOpen
  \bibfield  {author} {\bibinfo {author} {\bibfnamefont {J.~C.}\ \bibnamefont
  {Tully}},\ }\bibfield  {title} {\bibinfo {title} {Molecular dynamics with
  electronic transitions},\ }\href@noop {} {\bibfield  {journal} {\bibinfo
  {journal} {J. Chem. Phys.}\ }\textbf {\bibinfo {volume} {93}},\ \bibinfo
  {pages} {1061} (\bibinfo {year} {1990})}\BibitemShut {NoStop}%
\bibitem [{\citenamefont {Barbatti}(2011)}]{barbatti2011nonadiabatic}%
  \BibitemOpen
  \bibfield  {author} {\bibinfo {author} {\bibfnamefont {M.}~\bibnamefont
  {Barbatti}},\ }\bibfield  {title} {\bibinfo {title} {Nonadiabatic dynamics
  with trajectory surface hopping method},\ }\href@noop {} {\bibfield
  {journal} {\bibinfo  {journal} {WIREs Comput. Mol. Sci.}\ }\textbf {\bibinfo
  {volume} {1}},\ \bibinfo {pages} {620} (\bibinfo {year} {2011})}\BibitemShut
  {NoStop}%
\bibitem [{\citenamefont {Rebentrost}\ \emph {et~al.}(2009)\citenamefont
  {Rebentrost}, \citenamefont {Mohseni}, \citenamefont {Kassal}, \citenamefont
  {Lloyd},\ and\ \citenamefont {Aspuru-Guzik}}]{rebentrost2009environment}%
  \BibitemOpen
  \bibfield  {author} {\bibinfo {author} {\bibfnamefont {P.}~\bibnamefont
  {Rebentrost}}, \bibinfo {author} {\bibfnamefont {M.}~\bibnamefont {Mohseni}},
  \bibinfo {author} {\bibfnamefont {I.}~\bibnamefont {Kassal}}, \bibinfo
  {author} {\bibfnamefont {S.}~\bibnamefont {Lloyd}},\ and\ \bibinfo {author}
  {\bibfnamefont {A.}~\bibnamefont {Aspuru-Guzik}},\ }\bibfield  {title}
  {\bibinfo {title} {Environment-assisted quantum transport},\ }\href@noop {}
  {\bibfield  {journal} {\bibinfo  {journal} {New J. Phys.}\ }\textbf {\bibinfo
  {volume} {11}},\ \bibinfo {pages} {033003} (\bibinfo {year}
  {2009})}\BibitemShut {NoStop}%
\bibitem [{\citenamefont {Maier}\ \emph {et~al.}(2019)\citenamefont {Maier},
  \citenamefont {Brydges}, \citenamefont {Jurcevic}, \citenamefont {Trautmann},
  \citenamefont {Hempel}, \citenamefont {Lanyon}, \citenamefont {Hauke},
  \citenamefont {Blatt},\ and\ \citenamefont {Roos}}]{maier2019environment}%
  \BibitemOpen
  \bibfield  {author} {\bibinfo {author} {\bibfnamefont {C.}~\bibnamefont
  {Maier}}, \bibinfo {author} {\bibfnamefont {T.}~\bibnamefont {Brydges}},
  \bibinfo {author} {\bibfnamefont {P.}~\bibnamefont {Jurcevic}}, \bibinfo
  {author} {\bibfnamefont {N.}~\bibnamefont {Trautmann}}, \bibinfo {author}
  {\bibfnamefont {C.}~\bibnamefont {Hempel}}, \bibinfo {author} {\bibfnamefont
  {B.~P.}\ \bibnamefont {Lanyon}}, \bibinfo {author} {\bibfnamefont
  {P.}~\bibnamefont {Hauke}}, \bibinfo {author} {\bibfnamefont
  {R.}~\bibnamefont {Blatt}},\ and\ \bibinfo {author} {\bibfnamefont {C.~F.}\
  \bibnamefont {Roos}},\ }\bibfield  {title} {\bibinfo {title}
  {Environment-assisted quantum transport in a 10-qubit network},\ }\href@noop
  {} {\bibfield  {journal} {\bibinfo  {journal} {Phys. Rev. Lett.}\ }\textbf
  {\bibinfo {volume} {122}},\ \bibinfo {pages} {050501} (\bibinfo {year}
  {2019})}\BibitemShut {NoStop}%
\bibitem [{\citenamefont {Uys}\ \emph {et~al.}(2009)\citenamefont {Uys},
  \citenamefont {Biercuk},\ and\ \citenamefont {Bollinger}}]{uys2009optimized}%
  \BibitemOpen
  \bibfield  {author} {\bibinfo {author} {\bibfnamefont {H.}~\bibnamefont
  {Uys}}, \bibinfo {author} {\bibfnamefont {M.~J.}\ \bibnamefont {Biercuk}},\
  and\ \bibinfo {author} {\bibfnamefont {J.~J.}\ \bibnamefont {Bollinger}},\
  }\bibfield  {title} {\bibinfo {title} {Optimized noise filtration through
  dynamical decoupling},\ }\href@noop {} {\bibfield  {journal} {\bibinfo
  {journal} {Phys. Rev. Lett.}\ }\textbf {\bibinfo {volume} {103}},\ \bibinfo
  {pages} {040501} (\bibinfo {year} {2009})}\BibitemShut {NoStop}%
\bibitem [{\citenamefont {Lo~Franco}\ \emph {et~al.}(2014)\citenamefont
  {Lo~Franco}, \citenamefont {D'Arrigo}, \citenamefont {Falci}, \citenamefont
  {Compagno},\ and\ \citenamefont {Paladino}}]{lo2014preserving}%
  \BibitemOpen
  \bibfield  {author} {\bibinfo {author} {\bibfnamefont {R.}~\bibnamefont
  {Lo~Franco}}, \bibinfo {author} {\bibfnamefont {A.}~\bibnamefont {D'Arrigo}},
  \bibinfo {author} {\bibfnamefont {G.}~\bibnamefont {Falci}}, \bibinfo
  {author} {\bibfnamefont {G.}~\bibnamefont {Compagno}},\ and\ \bibinfo
  {author} {\bibfnamefont {E.}~\bibnamefont {Paladino}},\ }\bibfield  {title}
  {\bibinfo {title} {Preserving entanglement and nonlocality in solid-state
  qubits by dynamical decoupling},\ }\href@noop {} {\bibfield  {journal}
  {\bibinfo  {journal} {Phys. Rev. B}\ }\textbf {\bibinfo {volume} {90}},\
  \bibinfo {pages} {054304} (\bibinfo {year} {2014})}\BibitemShut {NoStop}%
\bibitem [{\citenamefont {Sza{\'n}kowski}\ \emph {et~al.}(2017)\citenamefont
  {Sza{\'n}kowski}, \citenamefont {Ramon}, \citenamefont {Krzywda},
  \citenamefont {Kwiatkowski} \emph {et~al.}}]{szankowski2017environmental}%
  \BibitemOpen
  \bibfield  {author} {\bibinfo {author} {\bibfnamefont {P.}~\bibnamefont
  {Sza{\'n}kowski}}, \bibinfo {author} {\bibfnamefont {G.}~\bibnamefont
  {Ramon}}, \bibinfo {author} {\bibfnamefont {J.}~\bibnamefont {Krzywda}},
  \bibinfo {author} {\bibfnamefont {D.}~\bibnamefont {Kwiatkowski}}, \emph
  {et~al.},\ }\bibfield  {title} {\bibinfo {title} {Environmental noise
  spectroscopy with qubits subjected to dynamical decoupling},\ }\href@noop {}
  {\bibfield  {journal} {\bibinfo  {journal} {J. Phys. Condens. Mat.}\ }\textbf
  {\bibinfo {volume} {29}},\ \bibinfo {pages} {333001} (\bibinfo {year}
  {2017})}\BibitemShut {NoStop}%
\bibitem [{\citenamefont {Brownnutt}\ \emph {et~al.}(2015)\citenamefont
  {Brownnutt}, \citenamefont {Kumph}, \citenamefont {Rabl},\ and\ \citenamefont
  {Blatt}}]{brownnutt2015ion}%
  \BibitemOpen
  \bibfield  {author} {\bibinfo {author} {\bibfnamefont {M.}~\bibnamefont
  {Brownnutt}}, \bibinfo {author} {\bibfnamefont {M.}~\bibnamefont {Kumph}},
  \bibinfo {author} {\bibfnamefont {P.}~\bibnamefont {Rabl}},\ and\ \bibinfo
  {author} {\bibfnamefont {R.}~\bibnamefont {Blatt}},\ }\bibfield  {title}
  {\bibinfo {title} {Ion-trap measurements of electric-field noise near
  surfaces},\ }\href@noop {} {\bibfield  {journal} {\bibinfo  {journal} {Rev.
  Mod. Phys.}\ }\textbf {\bibinfo {volume} {87}},\ \bibinfo {pages} {1419}
  (\bibinfo {year} {2015})}\BibitemShut {NoStop}%
\bibitem [{\citenamefont {Hite}\ \emph {et~al.}(2012)\citenamefont {Hite},
  \citenamefont {Colombe}, \citenamefont {Wilson}, \citenamefont {Brown},
  \citenamefont {Warring}, \citenamefont {J{\"o}rdens}, \citenamefont {Jost},
  \citenamefont {McKay}, \citenamefont {Pappas}, \citenamefont {Leibfried}
  \emph {et~al.}}]{hite2012100}%
  \BibitemOpen
  \bibfield  {author} {\bibinfo {author} {\bibfnamefont {D.~A.}\ \bibnamefont
  {Hite}}, \bibinfo {author} {\bibfnamefont {Y.}~\bibnamefont {Colombe}},
  \bibinfo {author} {\bibfnamefont {A.~C.}\ \bibnamefont {Wilson}}, \bibinfo
  {author} {\bibfnamefont {K.~R.}\ \bibnamefont {Brown}}, \bibinfo {author}
  {\bibfnamefont {U.}~\bibnamefont {Warring}}, \bibinfo {author} {\bibfnamefont
  {R.}~\bibnamefont {J{\"o}rdens}}, \bibinfo {author} {\bibfnamefont {J.~D.}\
  \bibnamefont {Jost}}, \bibinfo {author} {\bibfnamefont {K.}~\bibnamefont
  {McKay}}, \bibinfo {author} {\bibfnamefont {D.}~\bibnamefont {Pappas}},
  \bibinfo {author} {\bibfnamefont {D.}~\bibnamefont {Leibfried}}, \emph
  {et~al.},\ }\bibfield  {title} {\bibinfo {title} {100-fold reduction of
  electric-field noise in an ion trap cleaned with in situ argon-ion-beam
  bombardment},\ }\href@noop {} {\bibfield  {journal} {\bibinfo  {journal}
  {Phys. Rev. Lett.}\ }\textbf {\bibinfo {volume} {109}},\ \bibinfo {pages}
  {103001} (\bibinfo {year} {2012})}\BibitemShut {NoStop}%
\bibitem [{\citenamefont {Zeng}\ \emph {et~al.}(2023)\citenamefont {Zeng},
  \citenamefont {Yan}, \citenamefont {Yao}, \citenamefont {Zhao}, \citenamefont
  {Chen},\ and\ \citenamefont {Feng}}]{zeng2023theoretical}%
  \BibitemOpen
  \bibfield  {author} {\bibinfo {author} {\bibfnamefont {J.}~\bibnamefont
  {Zeng}}, \bibinfo {author} {\bibfnamefont {X.-T.}\ \bibnamefont {Yan}},
  \bibinfo {author} {\bibfnamefont {Y.}~\bibnamefont {Yao}}, \bibinfo {author}
  {\bibfnamefont {Y.-J.}\ \bibnamefont {Zhao}}, \bibinfo {author}
  {\bibfnamefont {L.}~\bibnamefont {Chen}},\ and\ \bibinfo {author}
  {\bibfnamefont {M.}~\bibnamefont {Feng}},\ }\bibfield  {title} {\bibinfo
  {title} {Theoretical investigation of electric-field noise emanating from
  vibrational adatoms},\ }\href@noop {} {\bibfield  {journal} {\bibinfo
  {journal} {Phys. Rev. A}\ }\textbf {\bibinfo {volume} {108}},\ \bibinfo
  {pages} {023121} (\bibinfo {year} {2023})}\BibitemShut {NoStop}%
\bibitem [{\citenamefont {Schmidt-Kaler}\ \emph {et~al.}(2003)\citenamefont
  {Schmidt-Kaler}, \citenamefont {Gulde}, \citenamefont {Riebe}, \citenamefont
  {Deuschle}, \citenamefont {Kreuter}, \citenamefont {Lancaster}, \citenamefont
  {Becher}, \citenamefont {Eschner}, \citenamefont {H{\"a}ffner},\ and\
  \citenamefont {Blatt}}]{schmidt2003coherence}%
  \BibitemOpen
  \bibfield  {author} {\bibinfo {author} {\bibfnamefont {F.}~\bibnamefont
  {Schmidt-Kaler}}, \bibinfo {author} {\bibfnamefont {S.}~\bibnamefont
  {Gulde}}, \bibinfo {author} {\bibfnamefont {M.}~\bibnamefont {Riebe}},
  \bibinfo {author} {\bibfnamefont {T.}~\bibnamefont {Deuschle}}, \bibinfo
  {author} {\bibfnamefont {A.}~\bibnamefont {Kreuter}}, \bibinfo {author}
  {\bibfnamefont {G.}~\bibnamefont {Lancaster}}, \bibinfo {author}
  {\bibfnamefont {C.}~\bibnamefont {Becher}}, \bibinfo {author} {\bibfnamefont
  {J.}~\bibnamefont {Eschner}}, \bibinfo {author} {\bibfnamefont
  {H.}~\bibnamefont {H{\"a}ffner}},\ and\ \bibinfo {author} {\bibfnamefont
  {R.}~\bibnamefont {Blatt}},\ }\bibfield  {title} {\bibinfo {title} {The
  coherence of qubits based on single {Ca$^+$} ions},\ }\href@noop {}
  {\bibfield  {journal} {\bibinfo  {journal} {J. Phys. B}\ }\textbf {\bibinfo
  {volume} {36}},\ \bibinfo {pages} {623} (\bibinfo {year} {2003})}\BibitemShut
  {NoStop}%
\bibitem [{\citenamefont {Merkel}\ \emph {et~al.}(2019)\citenamefont {Merkel},
  \citenamefont {Thirumalai}, \citenamefont {Tarlton}, \citenamefont
  {Sch{\"a}fer}, \citenamefont {Ballance}, \citenamefont {Harty},\ and\
  \citenamefont {Lucas}}]{merkel2019magnetic}%
  \BibitemOpen
  \bibfield  {author} {\bibinfo {author} {\bibfnamefont {B.}~\bibnamefont
  {Merkel}}, \bibinfo {author} {\bibfnamefont {K.}~\bibnamefont {Thirumalai}},
  \bibinfo {author} {\bibfnamefont {J.}~\bibnamefont {Tarlton}}, \bibinfo
  {author} {\bibfnamefont {V.}~\bibnamefont {Sch{\"a}fer}}, \bibinfo {author}
  {\bibfnamefont {C.}~\bibnamefont {Ballance}}, \bibinfo {author}
  {\bibfnamefont {T.}~\bibnamefont {Harty}},\ and\ \bibinfo {author}
  {\bibfnamefont {D.}~\bibnamefont {Lucas}},\ }\bibfield  {title} {\bibinfo
  {title} {Magnetic field stabilization system for atomic physics
  experiments},\ }\href@noop {} {\bibfield  {journal} {\bibinfo  {journal}
  {Rev. Sci. Instrum.}\ }\textbf {\bibinfo {volume} {90}} (\bibinfo {year}
  {2019})}\BibitemShut {NoStop}%
\bibitem [{\citenamefont {Viola}\ and\ \citenamefont
  {Lloyd}(1998)}]{viola1998dynamical}%
  \BibitemOpen
  \bibfield  {author} {\bibinfo {author} {\bibfnamefont {L.}~\bibnamefont
  {Viola}}\ and\ \bibinfo {author} {\bibfnamefont {S.}~\bibnamefont {Lloyd}},\
  }\bibfield  {title} {\bibinfo {title} {Dynamical suppression of decoherence
  in two-state quantum systems},\ }\href@noop {} {\bibfield  {journal}
  {\bibinfo  {journal} {Phys. Rev. A}\ }\textbf {\bibinfo {volume} {58}},\
  \bibinfo {pages} {2733} (\bibinfo {year} {1998})}\BibitemShut {NoStop}%
\bibitem [{\citenamefont {Viola}\ \emph {et~al.}(1999)\citenamefont {Viola},
  \citenamefont {Knill},\ and\ \citenamefont {Lloyd}}]{viola1999dynamical}%
  \BibitemOpen
  \bibfield  {author} {\bibinfo {author} {\bibfnamefont {L.}~\bibnamefont
  {Viola}}, \bibinfo {author} {\bibfnamefont {E.}~\bibnamefont {Knill}},\ and\
  \bibinfo {author} {\bibfnamefont {S.}~\bibnamefont {Lloyd}},\ }\bibfield
  {title} {\bibinfo {title} {Dynamical decoupling of open quantum systems},\
  }\href@noop {} {\bibfield  {journal} {\bibinfo  {journal} {Phys. Rev. Lett.}\
  }\textbf {\bibinfo {volume} {82}},\ \bibinfo {pages} {2417} (\bibinfo {year}
  {1999})}\BibitemShut {NoStop}%
\bibitem [{\citenamefont {Kiely}(2021)}]{kiely2021exact}%
  \BibitemOpen
  \bibfield  {author} {\bibinfo {author} {\bibfnamefont {A.}~\bibnamefont
  {Kiely}},\ }\bibfield  {title} {\bibinfo {title} {Exact classical noise
  master equations: Applications and connections},\ }\href@noop {} {\bibfield
  {journal} {\bibinfo  {journal} {EPL}\ }\textbf {\bibinfo {volume} {134}},\
  \bibinfo {pages} {10001} (\bibinfo {year} {2021})}\BibitemShut {NoStop}%
\bibitem [{\citenamefont {Deutsch}(1991)}]{deutsch1991quantum}%
  \BibitemOpen
  \bibfield  {author} {\bibinfo {author} {\bibfnamefont {J.~M.}\ \bibnamefont
  {Deutsch}},\ }\bibfield  {title} {\bibinfo {title} {Quantum statistical
  mechanics in a closed system},\ }\href@noop {} {\bibfield  {journal}
  {\bibinfo  {journal} {Phys. Rev. A}\ }\textbf {\bibinfo {volume} {43}},\
  \bibinfo {pages} {2046} (\bibinfo {year} {1991})}\BibitemShut {NoStop}%
\bibitem [{\citenamefont {Srednicki}(1994)}]{srednicki1994chaos}%
  \BibitemOpen
  \bibfield  {author} {\bibinfo {author} {\bibfnamefont {M.}~\bibnamefont
  {Srednicki}},\ }\bibfield  {title} {\bibinfo {title} {Chaos and quantum
  thermalization},\ }\href@noop {} {\bibfield  {journal} {\bibinfo  {journal}
  {Phys. Rev. E}\ }\textbf {\bibinfo {volume} {50}},\ \bibinfo {pages} {888}
  (\bibinfo {year} {1994})}\BibitemShut {NoStop}%
\bibitem [{\citenamefont {D'Alessio}\ \emph {et~al.}(2016)\citenamefont
  {D'Alessio}, \citenamefont {Kafri}, \citenamefont {Polkovnikov},\ and\
  \citenamefont {Rigol}}]{d2016quantum}%
  \BibitemOpen
  \bibfield  {author} {\bibinfo {author} {\bibfnamefont {L.}~\bibnamefont
  {D'Alessio}}, \bibinfo {author} {\bibfnamefont {Y.}~\bibnamefont {Kafri}},
  \bibinfo {author} {\bibfnamefont {A.}~\bibnamefont {Polkovnikov}},\ and\
  \bibinfo {author} {\bibfnamefont {M.}~\bibnamefont {Rigol}},\ }\bibfield
  {title} {\bibinfo {title} {From quantum chaos and eigenstate thermalization
  to statistical mechanics and thermodynamics},\ }\href@noop {} {\bibfield
  {journal} {\bibinfo  {journal} {Adv. Phys.}\ }\textbf {\bibinfo {volume}
  {65}},\ \bibinfo {pages} {239} (\bibinfo {year} {2016})}\BibitemShut
  {NoStop}%
\bibitem [{\citenamefont {Deutsch}(2018)}]{deutsch2018eigenstate}%
  \BibitemOpen
  \bibfield  {author} {\bibinfo {author} {\bibfnamefont {J.~M.}\ \bibnamefont
  {Deutsch}},\ }\bibfield  {title} {\bibinfo {title} {Eigenstate thermalization
  hypothesis},\ }\href@noop {} {\bibfield  {journal} {\bibinfo  {journal} {Rep.
  Prog. Phys.}\ }\textbf {\bibinfo {volume} {81}},\ \bibinfo {pages} {082001}
  (\bibinfo {year} {2018})}\BibitemShut {NoStop}%
\bibitem [{\citenamefont {Abanin}\ \emph {et~al.}(2019)\citenamefont {Abanin},
  \citenamefont {Altman}, \citenamefont {Bloch},\ and\ \citenamefont
  {Serbyn}}]{abanin2019colloquium}%
  \BibitemOpen
  \bibfield  {author} {\bibinfo {author} {\bibfnamefont {D.~A.}\ \bibnamefont
  {Abanin}}, \bibinfo {author} {\bibfnamefont {E.}~\bibnamefont {Altman}},
  \bibinfo {author} {\bibfnamefont {I.}~\bibnamefont {Bloch}},\ and\ \bibinfo
  {author} {\bibfnamefont {M.}~\bibnamefont {Serbyn}},\ }\bibfield  {title}
  {\bibinfo {title} {Colloquium: Many-body localization, thermalization, and
  entanglement},\ }\href@noop {} {\bibfield  {journal} {\bibinfo  {journal}
  {Rev. Mod. Phys.}\ }\textbf {\bibinfo {volume} {91}},\ \bibinfo {pages}
  {021001} (\bibinfo {year} {2019})}\BibitemShut {NoStop}%
\bibitem [{\citenamefont {Mostame}\ \emph {et~al.}(2012)\citenamefont
  {Mostame}, \citenamefont {Rebentrost}, \citenamefont {Eisfeld}, \citenamefont
  {Kerman}, \citenamefont {Tsomokos},\ and\ \citenamefont
  {Aspuru-Guzik}}]{mostame2012quantum}%
  \BibitemOpen
  \bibfield  {author} {\bibinfo {author} {\bibfnamefont {S.}~\bibnamefont
  {Mostame}}, \bibinfo {author} {\bibfnamefont {P.}~\bibnamefont {Rebentrost}},
  \bibinfo {author} {\bibfnamefont {A.}~\bibnamefont {Eisfeld}}, \bibinfo
  {author} {\bibfnamefont {A.~J.}\ \bibnamefont {Kerman}}, \bibinfo {author}
  {\bibfnamefont {D.~I.}\ \bibnamefont {Tsomokos}},\ and\ \bibinfo {author}
  {\bibfnamefont {A.}~\bibnamefont {Aspuru-Guzik}},\ }\bibfield  {title}
  {\bibinfo {title} {Quantum simulator of an open quantum system using
  superconducting qubits: exciton transport in photosynthetic complexes},\
  }\href@noop {} {\bibfield  {journal} {\bibinfo  {journal} {New J. Phys.}\
  }\textbf {\bibinfo {volume} {14}},\ \bibinfo {pages} {105013} (\bibinfo
  {year} {2012})}\BibitemShut {NoStop}%
\bibitem [{\citenamefont {Poto{\v{c}}nik}\ \emph {et~al.}(2018)\citenamefont
  {Poto{\v{c}}nik}, \citenamefont {Bargerbos}, \citenamefont {Schr{\"o}der},
  \citenamefont {Khan}, \citenamefont {Collodo}, \citenamefont {Gasparinetti},
  \citenamefont {Salath{\'e}}, \citenamefont {Creatore}, \citenamefont
  {Eichler}, \citenamefont {T{\"u}reci} \emph
  {et~al.}}]{potovcnik2018studying}%
  \BibitemOpen
  \bibfield  {author} {\bibinfo {author} {\bibfnamefont {A.}~\bibnamefont
  {Poto{\v{c}}nik}}, \bibinfo {author} {\bibfnamefont {A.}~\bibnamefont
  {Bargerbos}}, \bibinfo {author} {\bibfnamefont {F.~A.}\ \bibnamefont
  {Schr{\"o}der}}, \bibinfo {author} {\bibfnamefont {S.~A.}\ \bibnamefont
  {Khan}}, \bibinfo {author} {\bibfnamefont {M.~C.}\ \bibnamefont {Collodo}},
  \bibinfo {author} {\bibfnamefont {S.}~\bibnamefont {Gasparinetti}}, \bibinfo
  {author} {\bibfnamefont {Y.}~\bibnamefont {Salath{\'e}}}, \bibinfo {author}
  {\bibfnamefont {C.}~\bibnamefont {Creatore}}, \bibinfo {author}
  {\bibfnamefont {C.}~\bibnamefont {Eichler}}, \bibinfo {author} {\bibfnamefont
  {H.~E.}\ \bibnamefont {T{\"u}reci}}, \emph {et~al.},\ }\bibfield  {title}
  {\bibinfo {title} {Studying light-harvesting models with superconducting
  circuits},\ }\href@noop {} {\bibfield  {journal} {\bibinfo  {journal} {Nat.
  Commun.}\ }\textbf {\bibinfo {volume} {9}},\ \bibinfo {pages} {904} (\bibinfo
  {year} {2018})}\BibitemShut {NoStop}%
\bibitem [{\citenamefont {Wang}\ \emph {et~al.}(2018)\citenamefont {Wang},
  \citenamefont {Tao}, \citenamefont {Ai}, \citenamefont {Xin}, \citenamefont
  {Lambert}, \citenamefont {Ruan}, \citenamefont {Cheng}, \citenamefont {Nori},
  \citenamefont {Deng},\ and\ \citenamefont {Long}}]{wang2018efficient}%
  \BibitemOpen
  \bibfield  {author} {\bibinfo {author} {\bibfnamefont {B.-X.}\ \bibnamefont
  {Wang}}, \bibinfo {author} {\bibfnamefont {M.-J.}\ \bibnamefont {Tao}},
  \bibinfo {author} {\bibfnamefont {Q.}~\bibnamefont {Ai}}, \bibinfo {author}
  {\bibfnamefont {T.}~\bibnamefont {Xin}}, \bibinfo {author} {\bibfnamefont
  {N.}~\bibnamefont {Lambert}}, \bibinfo {author} {\bibfnamefont
  {D.}~\bibnamefont {Ruan}}, \bibinfo {author} {\bibfnamefont {Y.-C.}\
  \bibnamefont {Cheng}}, \bibinfo {author} {\bibfnamefont {F.}~\bibnamefont
  {Nori}}, \bibinfo {author} {\bibfnamefont {F.-G.}\ \bibnamefont {Deng}},\
  and\ \bibinfo {author} {\bibfnamefont {G.-L.}\ \bibnamefont {Long}},\
  }\bibfield  {title} {\bibinfo {title} {Efficient quantum simulation of
  photosynthetic light harvesting},\ }\href@noop {} {\bibfield  {journal}
  {\bibinfo  {journal} {NPJ Quantum Inform.}\ }\textbf {\bibinfo {volume}
  {4}},\ \bibinfo {pages} {52} (\bibinfo {year} {2018})}\BibitemShut {NoStop}%
\bibitem [{\citenamefont {Chenu}\ \emph {et~al.}(2017)\citenamefont {Chenu},
  \citenamefont {Beau}, \citenamefont {Cao},\ and\ \citenamefont {del
  Campo}}]{chenu2017quantum}%
  \BibitemOpen
  \bibfield  {author} {\bibinfo {author} {\bibfnamefont {A.}~\bibnamefont
  {Chenu}}, \bibinfo {author} {\bibfnamefont {M.}~\bibnamefont {Beau}},
  \bibinfo {author} {\bibfnamefont {J.}~\bibnamefont {Cao}},\ and\ \bibinfo
  {author} {\bibfnamefont {A.}~\bibnamefont {del Campo}},\ }\bibfield  {title}
  {\bibinfo {title} {Quantum simulation of generic many-body open system
  dynamics using classical noise},\ }\href@noop {} {\bibfield  {journal}
  {\bibinfo  {journal} {Phys. Rev. Lett.}\ }\textbf {\bibinfo {volume} {118}},\
  \bibinfo {pages} {140403} (\bibinfo {year} {2017})}\BibitemShut {NoStop}%
\bibitem [{\citenamefont {Gu}\ and\ \citenamefont {Franco}(2019)}]{gu2019can}%
  \BibitemOpen
  \bibfield  {author} {\bibinfo {author} {\bibfnamefont {B.}~\bibnamefont
  {Gu}}\ and\ \bibinfo {author} {\bibfnamefont {I.}~\bibnamefont {Franco}},\
  }\bibfield  {title} {\bibinfo {title} {When can quantum decoherence be
  mimicked by classical noise?},\ }\href@noop {} {\bibfield  {journal}
  {\bibinfo  {journal} {J. Chem. Phys.}\ }\textbf {\bibinfo {volume} {151}},\
  \bibinfo {pages} {014109} (\bibinfo {year} {2019})}\BibitemShut {NoStop}%
\bibitem [{\citenamefont {Bergli}\ \emph {et~al.}(2009)\citenamefont {Bergli},
  \citenamefont {Galperin},\ and\ \citenamefont
  {Altshuler}}]{bergli2009decoherence}%
  \BibitemOpen
  \bibfield  {author} {\bibinfo {author} {\bibfnamefont {J.}~\bibnamefont
  {Bergli}}, \bibinfo {author} {\bibfnamefont {Y.~M.}\ \bibnamefont
  {Galperin}},\ and\ \bibinfo {author} {\bibfnamefont {B.}~\bibnamefont
  {Altshuler}},\ }\bibfield  {title} {\bibinfo {title} {Decoherence in qubits
  due to low-frequency noise},\ }\href@noop {} {\bibfield  {journal} {\bibinfo
  {journal} {New J. Phys.}\ }\textbf {\bibinfo {volume} {11}},\ \bibinfo
  {pages} {025002} (\bibinfo {year} {2009})}\BibitemShut {NoStop}%
\bibitem [{\citenamefont {Yang}\ \emph {et~al.}(2016)\citenamefont {Yang},
  \citenamefont {Ma},\ and\ \citenamefont {Liu}}]{yang2016quantum}%
  \BibitemOpen
  \bibfield  {author} {\bibinfo {author} {\bibfnamefont {W.}~\bibnamefont
  {Yang}}, \bibinfo {author} {\bibfnamefont {W.-L.}\ \bibnamefont {Ma}},\ and\
  \bibinfo {author} {\bibfnamefont {R.-B.}\ \bibnamefont {Liu}},\ }\bibfield
  {title} {\bibinfo {title} {Quantum many-body theory for electron spin
  decoherence in nanoscale nuclear spin baths},\ }\href@noop {} {\bibfield
  {journal} {\bibinfo  {journal} {Rep. Prog. Phys.}\ }\textbf {\bibinfo
  {volume} {80}},\ \bibinfo {pages} {016001} (\bibinfo {year}
  {2016})}\BibitemShut {NoStop}%
\bibitem [{\citenamefont {Tanimura}\ and\ \citenamefont
  {Kubo}(1989)}]{tanimura1989time}%
  \BibitemOpen
  \bibfield  {author} {\bibinfo {author} {\bibfnamefont {Y.}~\bibnamefont
  {Tanimura}}\ and\ \bibinfo {author} {\bibfnamefont {R.}~\bibnamefont
  {Kubo}},\ }\bibfield  {title} {\bibinfo {title} {Time evolution of a quantum
  system in contact with a nearly {Gaussian-Markoffian} noise bath},\
  }\href@noop {} {\bibfield  {journal} {\bibinfo  {journal} {J. Phys. Soc.
  Jpn.}\ }\textbf {\bibinfo {volume} {58}},\ \bibinfo {pages} {101} (\bibinfo
  {year} {1989})}\BibitemShut {NoStop}%
\bibitem [{\citenamefont {Burgarth}\ \emph {et~al.}(2017)\citenamefont
  {Burgarth}, \citenamefont {Facchi}, \citenamefont {Garnero}, \citenamefont
  {Nakazato}, \citenamefont {Pascazio},\ and\ \citenamefont
  {Yuasa}}]{burgarth2017can}%
  \BibitemOpen
  \bibfield  {author} {\bibinfo {author} {\bibfnamefont {D.}~\bibnamefont
  {Burgarth}}, \bibinfo {author} {\bibfnamefont {P.}~\bibnamefont {Facchi}},
  \bibinfo {author} {\bibfnamefont {G.}~\bibnamefont {Garnero}}, \bibinfo
  {author} {\bibfnamefont {H.}~\bibnamefont {Nakazato}}, \bibinfo {author}
  {\bibfnamefont {S.}~\bibnamefont {Pascazio}},\ and\ \bibinfo {author}
  {\bibfnamefont {K.}~\bibnamefont {Yuasa}},\ }\bibfield  {title} {\bibinfo
  {title} {Can decay be ascribed to classical noise?},\ }\href@noop {}
  {\bibfield  {journal} {\bibinfo  {journal} {Open Syst. Inf. Dyn.}\ }\textbf
  {\bibinfo {volume} {24}},\ \bibinfo {pages} {1750001} (\bibinfo {year}
  {2017})}\BibitemShut {NoStop}%
\bibitem [{\citenamefont {Haga}\ \emph {et~al.}(2023)\citenamefont {Haga},
  \citenamefont {Nakagawa}, \citenamefont {Hamazaki},\ and\ \citenamefont
  {Ueda}}]{haga2023quasiparticles}%
  \BibitemOpen
  \bibfield  {author} {\bibinfo {author} {\bibfnamefont {T.}~\bibnamefont
  {Haga}}, \bibinfo {author} {\bibfnamefont {M.}~\bibnamefont {Nakagawa}},
  \bibinfo {author} {\bibfnamefont {R.}~\bibnamefont {Hamazaki}},\ and\
  \bibinfo {author} {\bibfnamefont {M.}~\bibnamefont {Ueda}},\ }\bibfield
  {title} {\bibinfo {title} {Quasiparticles of decoherence processes in open
  quantum many-body systems: Incoherentons},\ }\href@noop {} {\bibfield
  {journal} {\bibinfo  {journal} {Phys. Rev. Res.}\ }\textbf {\bibinfo {volume}
  {5}},\ \bibinfo {pages} {043225} (\bibinfo {year} {2023})}\BibitemShut
  {NoStop}%
\bibitem [{\citenamefont {Bastida}\ \emph {et~al.}(2007)\citenamefont
  {Bastida}, \citenamefont {Cruz}, \citenamefont {Z{\'u}{\~n}iga},
  \citenamefont {Requena},\ and\ \citenamefont
  {Miguel}}]{bastida2007ehrenfest}%
  \BibitemOpen
  \bibfield  {author} {\bibinfo {author} {\bibfnamefont {A.}~\bibnamefont
  {Bastida}}, \bibinfo {author} {\bibfnamefont {C.}~\bibnamefont {Cruz}},
  \bibinfo {author} {\bibfnamefont {J.}~\bibnamefont {Z{\'u}{\~n}iga}},
  \bibinfo {author} {\bibfnamefont {A.}~\bibnamefont {Requena}},\ and\ \bibinfo
  {author} {\bibfnamefont {B.}~\bibnamefont {Miguel}},\ }\bibfield  {title}
  {\bibinfo {title} {The ehrenfest method with quantum corrections to simulate
  the relaxation of molecules in solution: Equilibrium and dynamics},\
  }\href@noop {} {\bibfield  {journal} {\bibinfo  {journal} {J. Chem. Phys.}\
  }\textbf {\bibinfo {volume} {126}},\ \bibinfo {pages} {014503} (\bibinfo
  {year} {2007})}\BibitemShut {NoStop}%
\bibitem [{\citenamefont {Aghtar}\ \emph {et~al.}(2012)\citenamefont {Aghtar},
  \citenamefont {Liebers}, \citenamefont {Str{\"u}mpfer}, \citenamefont
  {Schulten},\ and\ \citenamefont
  {Kleinekath{\"o}fer}}]{aghtar2012juxtaposing}%
  \BibitemOpen
  \bibfield  {author} {\bibinfo {author} {\bibfnamefont {M.}~\bibnamefont
  {Aghtar}}, \bibinfo {author} {\bibfnamefont {J.}~\bibnamefont {Liebers}},
  \bibinfo {author} {\bibfnamefont {J.}~\bibnamefont {Str{\"u}mpfer}}, \bibinfo
  {author} {\bibfnamefont {K.}~\bibnamefont {Schulten}},\ and\ \bibinfo
  {author} {\bibfnamefont {U.}~\bibnamefont {Kleinekath{\"o}fer}},\ }\bibfield
  {title} {\bibinfo {title} {Juxtaposing density matrix and classical
  path-based wave packet dynamics},\ }\href@noop {} {\bibfield  {journal}
  {\bibinfo  {journal} {J. Chem. Phys.}\ }\textbf {\bibinfo {volume} {136}},\
  \bibinfo {pages} {214101} (\bibinfo {year} {2012})}\BibitemShut {NoStop}%
\bibitem [{\citenamefont {Lindblad}(1976)}]{lindblad1976generators}%
  \BibitemOpen
  \bibfield  {author} {\bibinfo {author} {\bibfnamefont {G.}~\bibnamefont
  {Lindblad}},\ }\bibfield  {title} {\bibinfo {title} {On the generators of
  quantum dynamical semigroups},\ }\href@noop {} {\bibfield  {journal}
  {\bibinfo  {journal} {Commun. Math. Phys.}\ }\textbf {\bibinfo {volume}
  {48}},\ \bibinfo {pages} {119} (\bibinfo {year} {1976})}\BibitemShut
  {NoStop}%
\bibitem [{\citenamefont {Manzano}(2020)}]{manzano2020short}%
  \BibitemOpen
  \bibfield  {author} {\bibinfo {author} {\bibfnamefont {D.}~\bibnamefont
  {Manzano}},\ }\bibfield  {title} {\bibinfo {title} {A short introduction to
  the {Lindblad} master equation},\ }\href@noop {} {\bibfield  {journal}
  {\bibinfo  {journal} {AIP Adv.}\ }\textbf {\bibinfo {volume} {10}},\ \bibinfo
  {pages} {025106} (\bibinfo {year} {2020})}\BibitemShut {NoStop}%
\bibitem [{\citenamefont {Breuer}\ and\ \citenamefont
  {Petruccione}(2002)}]{breuer2002theory}%
  \BibitemOpen
  \bibfield  {author} {\bibinfo {author} {\bibfnamefont {H.-P.}\ \bibnamefont
  {Breuer}}\ and\ \bibinfo {author} {\bibfnamefont {F.}~\bibnamefont
  {Petruccione}},\ }\href@noop {} {\emph {\bibinfo {title} {The theory of open
  quantum systems}}}\ (\bibinfo  {publisher} {Oxford University Press},\
  \bibinfo {year} {2002})\BibitemShut {NoStop}%
\bibitem [{\citenamefont {Costa-Filho}\ \emph {et~al.}(2017)\citenamefont
  {Costa-Filho}, \citenamefont {Lima}, \citenamefont {Paiva}, \citenamefont
  {Soares}, \citenamefont {Morgado}, \citenamefont {Franco},\ and\
  \citenamefont {Soares-Pinto}}]{costa2017enabling}%
  \BibitemOpen
  \bibfield  {author} {\bibinfo {author} {\bibfnamefont {J.}~\bibnamefont
  {Costa-Filho}}, \bibinfo {author} {\bibfnamefont {R.}~\bibnamefont {Lima}},
  \bibinfo {author} {\bibfnamefont {R.}~\bibnamefont {Paiva}}, \bibinfo
  {author} {\bibfnamefont {P.}~\bibnamefont {Soares}}, \bibinfo {author}
  {\bibfnamefont {W.}~\bibnamefont {Morgado}}, \bibinfo {author} {\bibfnamefont
  {R.~L.}\ \bibnamefont {Franco}},\ and\ \bibinfo {author} {\bibfnamefont
  {D.}~\bibnamefont {Soares-Pinto}},\ }\bibfield  {title} {\bibinfo {title}
  {Enabling quantum non-{Markovian} dynamics by injection of classical colored
  noise},\ }\href@noop {} {\bibfield  {journal} {\bibinfo  {journal} {Phys.
  Rev. A}\ }\textbf {\bibinfo {volume} {95}},\ \bibinfo {pages} {052126}
  (\bibinfo {year} {2017})}\BibitemShut {NoStop}%
\bibitem [{\citenamefont {Breuer}(2012)}]{breuer2012foundations}%
  \BibitemOpen
  \bibfield  {author} {\bibinfo {author} {\bibfnamefont {H.-P.}\ \bibnamefont
  {Breuer}},\ }\bibfield  {title} {\bibinfo {title} {Foundations and measures
  of quantum non-{Markovianity}},\ }\href@noop {} {\bibfield  {journal}
  {\bibinfo  {journal} {J. Phys. B}\ }\textbf {\bibinfo {volume} {45}},\
  \bibinfo {pages} {154001} (\bibinfo {year} {2012})}\BibitemShut {NoStop}%
\bibitem [{\citenamefont {Wenderoth}\ \emph {et~al.}(2021)\citenamefont
  {Wenderoth}, \citenamefont {Breuer},\ and\ \citenamefont
  {Thoss}}]{wenderoth2021non}%
  \BibitemOpen
  \bibfield  {author} {\bibinfo {author} {\bibfnamefont {S.}~\bibnamefont
  {Wenderoth}}, \bibinfo {author} {\bibfnamefont {H.-P.}\ \bibnamefont
  {Breuer}},\ and\ \bibinfo {author} {\bibfnamefont {M.}~\bibnamefont
  {Thoss}},\ }\bibfield  {title} {\bibinfo {title} {Non-{Markovian} effects in
  the spin-boson model at zero temperature},\ }\href@noop {} {\bibfield
  {journal} {\bibinfo  {journal} {Phys. Rev. A}\ }\textbf {\bibinfo {volume}
  {104}},\ \bibinfo {pages} {012213} (\bibinfo {year} {2021})}\BibitemShut
  {NoStop}%
\bibitem [{\citenamefont {Zeng}\ and\ \citenamefont
  {Yao}(2024)}]{zeng2024perturbative}%
  \BibitemOpen
  \bibfield  {author} {\bibinfo {author} {\bibfnamefont {J.}~\bibnamefont
  {Zeng}}\ and\ \bibinfo {author} {\bibfnamefont {Y.}~\bibnamefont {Yao}},\
  }\bibfield  {title} {\bibinfo {title} {Perturbative dynamics in the
  pseudocoherent phase of the spin-boson model},\ }\href@noop {} {\bibfield
  {journal} {\bibinfo  {journal} {Phys. Rev. A}\ }\textbf {\bibinfo {volume}
  {110}},\ \bibinfo {pages} {012212} (\bibinfo {year} {2024})}\BibitemShut
  {NoStop}%
\bibitem [{\citenamefont {Purcell}\ and\ \citenamefont
  {Pound}(1951)}]{purcell1951nuclear}%
  \BibitemOpen
  \bibfield  {author} {\bibinfo {author} {\bibfnamefont {E.~M.}\ \bibnamefont
  {Purcell}}\ and\ \bibinfo {author} {\bibfnamefont {R.~V.}\ \bibnamefont
  {Pound}},\ }\bibfield  {title} {\bibinfo {title} {A nuclear spin system at
  negative temperature},\ }\href@noop {} {\bibfield  {journal} {\bibinfo
  {journal} {Phys. Rev.}\ }\textbf {\bibinfo {volume} {81}},\ \bibinfo {pages}
  {279} (\bibinfo {year} {1951})}\BibitemShut {NoStop}%
\bibitem [{\citenamefont {Ramsey}(1956)}]{ramsey1956thermodynamics}%
  \BibitemOpen
  \bibfield  {author} {\bibinfo {author} {\bibfnamefont {N.~F.}\ \bibnamefont
  {Ramsey}},\ }\bibfield  {title} {\bibinfo {title} {Thermodynamics and
  statistical mechanics at negative absolute temperatures},\ }\href@noop {}
  {\bibfield  {journal} {\bibinfo  {journal} {Phys. Rev.}\ }\textbf {\bibinfo
  {volume} {103}},\ \bibinfo {pages} {20} (\bibinfo {year} {1956})}\BibitemShut
  {NoStop}%
\bibitem [{\citenamefont {Lapolla}\ and\ \citenamefont
  {Godec}(2020)}]{lapolla2020faster}%
  \BibitemOpen
  \bibfield  {author} {\bibinfo {author} {\bibfnamefont {A.}~\bibnamefont
  {Lapolla}}\ and\ \bibinfo {author} {\bibfnamefont {A.}~\bibnamefont
  {Godec}},\ }\bibfield  {title} {\bibinfo {title} {Faster uphill relaxation in
  thermodynamically equidistant temperature quenches},\ }\href@noop {}
  {\bibfield  {journal} {\bibinfo  {journal} {Phys. Rev. Lett.}\ }\textbf
  {\bibinfo {volume} {125}},\ \bibinfo {pages} {110602} (\bibinfo {year}
  {2020})}\BibitemShut {NoStop}%
\bibitem [{\citenamefont {Manikandan}(2021)}]{manikandan2021equidistant}%
  \BibitemOpen
  \bibfield  {author} {\bibinfo {author} {\bibfnamefont {S.~K.}\ \bibnamefont
  {Manikandan}},\ }\bibfield  {title} {\bibinfo {title} {Equidistant quenches
  in few-level quantum systems},\ }\href@noop {} {\bibfield  {journal}
  {\bibinfo  {journal} {Phys. Rev. Res.}\ }\textbf {\bibinfo {volume} {3}},\
  \bibinfo {pages} {043108} (\bibinfo {year} {2021})}\BibitemShut {NoStop}%
\bibitem [{\citenamefont {Bera}\ \emph {et~al.}(2024)\citenamefont {Bera},
  \citenamefont {Pandit}, \citenamefont {Chatterjee}, \citenamefont {Singh},
  \citenamefont {Lewenstein}, \citenamefont {Bhattacharya},\ and\ \citenamefont
  {Bera}}]{bera2024steady}%
  \BibitemOpen
  \bibfield  {author} {\bibinfo {author} {\bibfnamefont {M.~L.}\ \bibnamefont
  {Bera}}, \bibinfo {author} {\bibfnamefont {T.}~\bibnamefont {Pandit}},
  \bibinfo {author} {\bibfnamefont {K.}~\bibnamefont {Chatterjee}}, \bibinfo
  {author} {\bibfnamefont {V.}~\bibnamefont {Singh}}, \bibinfo {author}
  {\bibfnamefont {M.}~\bibnamefont {Lewenstein}}, \bibinfo {author}
  {\bibfnamefont {U.}~\bibnamefont {Bhattacharya}},\ and\ \bibinfo {author}
  {\bibfnamefont {M.~N.}\ \bibnamefont {Bera}},\ }\bibfield  {title} {\bibinfo
  {title} {Steady-state quantum thermodynamics with synthetic negative
  temperatures},\ }\href@noop {} {\bibfield  {journal} {\bibinfo  {journal}
  {Phys. Rev. Res.}\ }\textbf {\bibinfo {volume} {6}},\ \bibinfo {pages}
  {013318} (\bibinfo {year} {2024})}\BibitemShut {NoStop}%
\bibitem [{\citenamefont {Chen}\ and\ \citenamefont
  {Reichman}(2016)}]{chen2016accuracy}%
  \BibitemOpen
  \bibfield  {author} {\bibinfo {author} {\bibfnamefont {H.-T.}\ \bibnamefont
  {Chen}}\ and\ \bibinfo {author} {\bibfnamefont {D.~R.}\ \bibnamefont
  {Reichman}},\ }\bibfield  {title} {\bibinfo {title} {On the accuracy of
  surface hopping dynamics in condensed phase non-adiabatic problems},\
  }\href@noop {} {\bibfield  {journal} {\bibinfo  {journal} {J. Chem. Phys.}\
  }\textbf {\bibinfo {volume} {144}},\ \bibinfo {pages} {094104} (\bibinfo
  {year} {2016})}\BibitemShut {NoStop}%
\bibitem [{\citenamefont {Zagoskin}\ and\ \citenamefont
  {Blais}(2008)}]{zagoskin2008superconducting}%
  \BibitemOpen
  \bibfield  {author} {\bibinfo {author} {\bibfnamefont {A.}~\bibnamefont
  {Zagoskin}}\ and\ \bibinfo {author} {\bibfnamefont {A.}~\bibnamefont
  {Blais}},\ }\bibfield  {title} {\bibinfo {title} {Superconducting qubits},\
  }\href@noop {} {\bibfield  {journal} {\bibinfo  {journal} {arXiv: 0805.0164}\
  } (\bibinfo {year} {2008})}\BibitemShut {NoStop}%
\bibitem [{\citenamefont
  {Gillespie}(1996{\natexlab{a}})}]{gillespie1996mathematics}%
  \BibitemOpen
  \bibfield  {author} {\bibinfo {author} {\bibfnamefont {D.~T.}\ \bibnamefont
  {Gillespie}},\ }\bibfield  {title} {\bibinfo {title} {The mathematics of
  {Brownian} motion and {Johnson} noise},\ }\href@noop {} {\bibfield  {journal}
  {\bibinfo  {journal} {Am. J. Phys.}\ }\textbf {\bibinfo {volume} {64}},\
  \bibinfo {pages} {225} (\bibinfo {year} {1996}{\natexlab{a}})}\BibitemShut
  {NoStop}%
\bibitem [{\citenamefont {Gillespie}(1996{\natexlab{b}})}]{gillespie1996exact}%
  \BibitemOpen
  \bibfield  {author} {\bibinfo {author} {\bibfnamefont {D.~T.}\ \bibnamefont
  {Gillespie}},\ }\bibfield  {title} {\bibinfo {title} {Exact numerical
  simulation of the {Ornstein-Uhlenbeck} process and its integral},\
  }\href@noop {} {\bibfield  {journal} {\bibinfo  {journal} {Phys. Rev. E}\
  }\textbf {\bibinfo {volume} {54}},\ \bibinfo {pages} {2084} (\bibinfo {year}
  {1996}{\natexlab{b}})}\BibitemShut {NoStop}%
\end{thebibliography}%

\end{document}